\documentclass[letter]{aa}
\usepackage{graphicx}
\usepackage[varg]{txfonts}
\usepackage{color}
\begin{document}
\title{First imaging spectroscopy observations of solar drift pair bursts}
\author{A.A. Kuznetsov\inst{\ref{ISZF}}
        \and
        E.P. Kontar\inst{\ref{UoG}}}
\institute{Institute of Solar-Terrestrial Physics, Irkutsk 664033, Russia\label{ISZF}
           \and
           School of Physics and Astronomy, University of Glasgow, Glasgow G12 8QQ, UK\label{UoG}}
\titlerunning{First imaging spectroscopy observations of solar drift pair bursts}
\authorrunning{Kuznetsov and Kontar}
\date{Received *; accepted *}
\abstract{Drift pairs are an unusual and puzzling type of fine structure sometimes observed in dynamic spectra of solar radio emission. They appear as two identical short narrowband drifting stripes separated in time; both positive and negative frequency drifts are observed. Currently, due to the lack of imaging observations, there is no satisfactory explanation for this phenomenon. Using the Low Frequency Array (LOFAR), we report unique observations of a cluster of drift pair bursts in the frequency range of $30-70$ MHz made on 12 July 2017. Spectral imaging capabilities of the instrument have allowed us for the first time to resolve the temporal and frequency evolution of the source locations and sizes at a fixed frequency and along the drifting pair components. Sources of two components of a drift pair have been imaged and found to propagate in the same direction along nearly the same trajectories. Motion of the second component source is seen to be delayed in time with respect to that of the first one. The source trajectories can be complicated and non-radial; positive and negative frequency drifts correspond to opposite propagation directions. The drift pair bursts with positive and negative frequency drifts, as well as the associated broadband type-III-like bursts, are produced in the same regions. The visible source velocities are variable from zero to a few $10^4$ (up to ${\sim 10^5}$) km $\textrm{s}^{-1}$, which often exceeds the  velocities inferred from the drift rate ($\sim 10^4$ km $\textrm{s}^{-1}$). The visible source sizes are of about $10'-18'$; they are more compact than typical type III sources at the same frequencies. The existing models of drift pair bursts cannot adequately explain the observed features. We discuss the key issues that need to be addressed, and in particular the anisotropic scattering of the radio waves. The broadband bursts observed simultaneously with the drift pairs differ in some aspects from common type III bursts and may represent a separate type of emission.}

\maketitle

\section{Introduction}
Drift pair bursts (DPBs) were identified as a separate class of fine structures in the dynamic spectra of solar radio emission by \cite{rob58}. They occur at low frequencies ($\sim 10-100$ MHz) and appear as two parallel frequency-drifting narrow-band stripes separated in time. The drift rates are typically of about $2-8$ MHz $\textrm{s}^{-1}$, that is, they are intermediate between the drift rates of type II and type III bursts at the same frequencies; both positive and negative frequency drifts are observed. The delay of the second component is typically of about $1-2$~s. The DPBs are often associated with storms of type III bursts.

The most enigmatic characteristic of DPBs is that in contrast to other multi-stripe bursts (e.g., type II bursts and zebra patterns), their components are shifted in time rather than in frequency -- the second component of a pair looks like a repetition of the first one (although sometimes with a slightly different intensity), while the frequency shift is absent or very small \citep{ell69, noe71, mol78, mel82, mel05}. The polarization degree varies from very low to $\sim 50\%$; the higher-frequency component (i.e., the first component of the pairs with positive frequency drift and the second component of the pairs with negative frequency drift) generally has a higher polarization degree \citep{suz79, dul84}. Sometimes, unusual variations of DPBs are observed, such as the so-called hooks and structures with a third component \citep{ell69, mel05}.

The only published study of source positions and sizes of DPBs is by \citet{suz79}, who managed to determine positions for both pair components in three instances, using the Culgoora radioheliograph. These observations suggested that the source positions of two components of a pair coincided (within the instrument beam half-width of $\sim 4'$). We note, however, that these observations were performed at a single frequency (43 MHz), and with 3 s time cadence, which is evidently insufficient to resolve the dynamics of DPBs with typical time separations of $\lesssim 2$ s. The origin of the DPBs remains unclear; among a number of proposed theoretical models \citep[see, e.g., the reviews and discussions in the papers of][]{mel82, mel05}, no single model can adequately explain all the observed features. One of the reasons for this situation is the above-mentioned deficiency in the spatially resolved observations.

In this Letter, we present the first detailed imaging observations of DPBs in a broad frequency range made with the \textit{Low Frequency Array} (LOFAR). We do not aim to perform a complete statistical analysis and instead highlight the main features discovered.

\begin{figure*}
\includegraphics{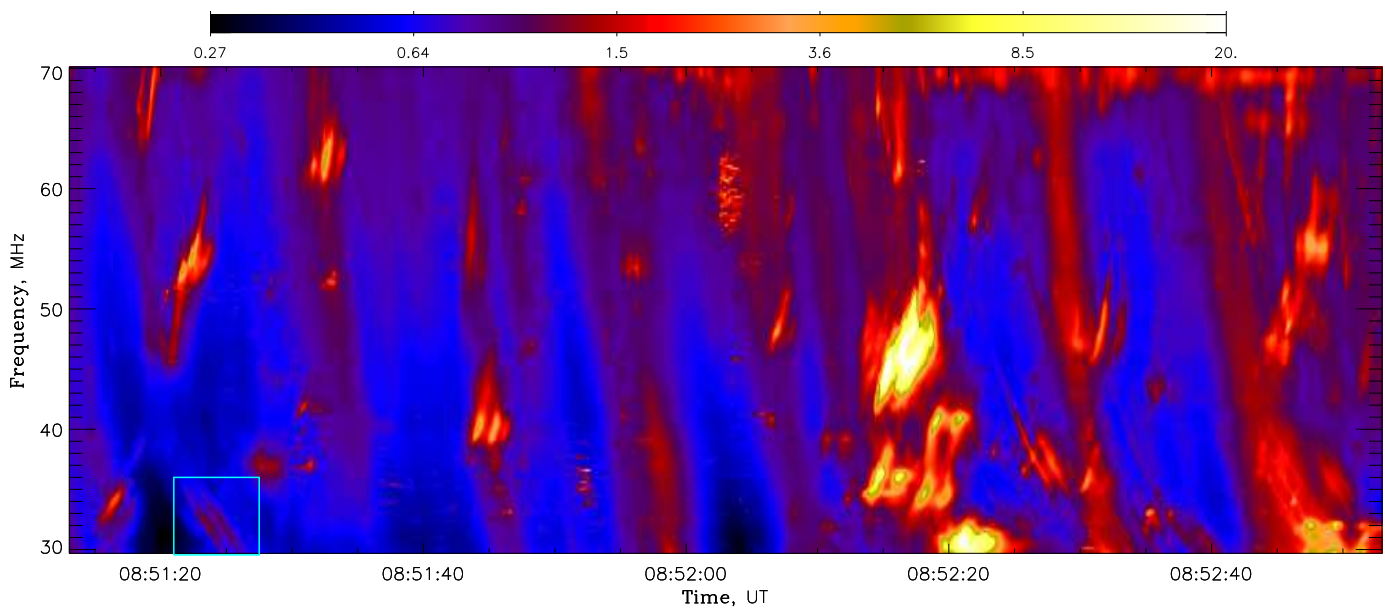}
\caption{Dynamic spectrum of solar radio emission recorded with LOFAR on 12 July 2017 in relative units (the radio flux normalized by the average flux in a given frequency channel).}
\label{DSlarge}
\end{figure*}

\section{Observations}
A large cluster of DPBs and broadband type-III-like bursts was recorded with the LOFAR \citep[see][]{haa13} on 12 July 2017, 08:39:00 -- 08:53:00 UT; the session duration was 14 minutes so that it covered only a part of the burst storm. The radio bursts were seemingly not associated with any other activity: no X-ray or optical flares were detected during the mentioned period or immediately before it.

The observations were made in the low-frequency band ($30-70$ MHz) using the LOFAR core stations. The imaging capabilities were achieved using the beam-formed mode when the data from the LOFAR antennae are combined to form tied-array beams covering an area of the sky; the beam size is of about $10'$ at 32 MHz. The used configuration included 217 beams with a mosaic beam spacing of about $6'$ to ensure partial overlapping \citep{kon17}; the radio fluxes were analyzed with the time and frequency resolutions of 0.1 s and 12 kHz, respectively (in the below examples, the data were sampled to the frequency resolution of 195 kHz). Unfortunately, the LOFAR polarization measurements were unavailable for this event.

\subsection{Dynamic spectra}
The total (spatially unresolved) dynamic spectra (Figure \ref{DSlarge}) are obtained by summing the LOFAR data over all beams. Figure \ref{DSlarge} shows an example of the spatially integrated dynamic radio spectrum containing a number of fine structures. Drift pair bursts with both positive and negative frequency drifts can be seen; the bursts with opposite drift signs can occur at the same times and frequencies and sometimes even intersect each other. Some of the DPBs appear superimposed on broadband drifting structures similar to type III bursts.

\begin{figure}
\includegraphics{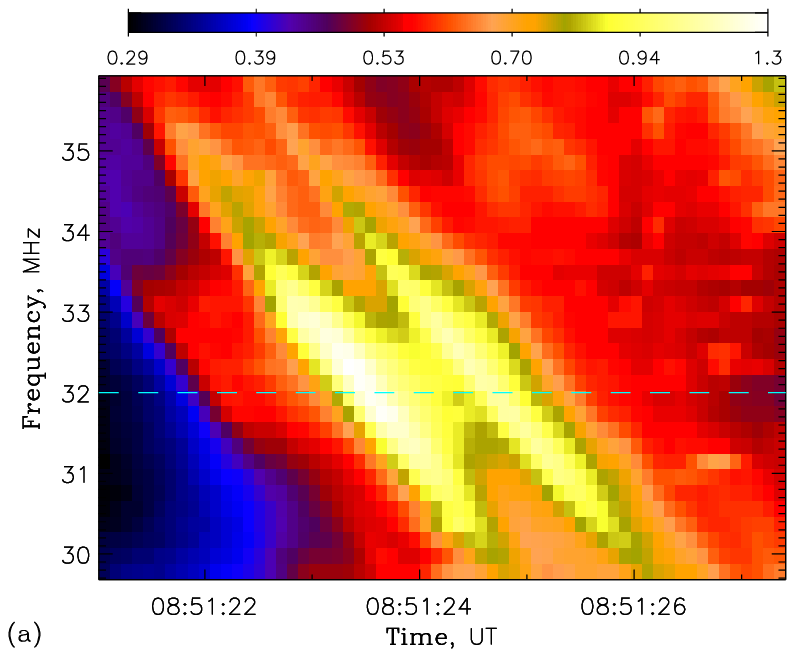}

\includegraphics{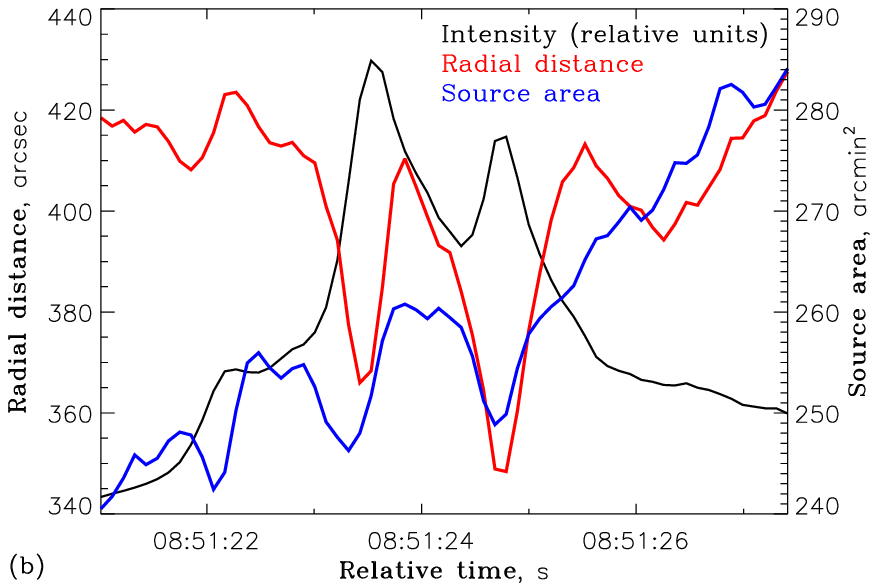}
\caption{Top: Enlarged fragment of dynamic spectrum with one DPB from Figure \protect\ref{DSlarge}. Bottom: Time profiles of the emission source parameters at a single frequency (32 MHz); black curve shows variations of the total intensity (units are not shown).}
\label{DSsmall}
\end{figure}

Figure \ref{DSsmall}a shows an enlarged fragment of dynamic spectrum with one DPB. This example demonstrates typical features of DPBs: the emission stripes are strictly parallel and appear shifted in time (by $\sim 1.5$ s); the stripes themselves are short ($\sim 0.7$ s) and narrowband ($\sim 1.5$ MHz).

All observed DPBs appear qualitatively similar, although their parameters may vary: for example, the frequency drift rates vary in the range of $\sim 1.5-6$ MHz $\textrm{s}^{-1}$; the drift rates increase with frequency.  Assuming the plasma emission and the \citet{new61} coronal density model, we estimate the corresponding exciter speed as $\sim 20\,000-25\,000$ km $\textrm{s}^{-1}$ -- a conclusion that is not supported by imaging observations (see below). The delays between the components of a pair are of about $1-2$ s. The DPBs with positive frequency drift are usually more diffuse (i.e., have a larger instantaneous bandwidth and a longer duration at a fixed frequency) than the bursts with negative frequency drift. Sometimes, bursts with three or more parallel stripes are observed, but these could  be formed due to alignment and/or overlapping of simpler bursts.

The broadband bursts have durations (at a fixed frequency) of about $2-5$ s and the negative frequency drift rates varying from $\sim 5$ to $\sim 30$ MHz~$\textrm{s}^{-1}$ in the frequency range of $30-70$ MHz. We note that the drift rates at lower frequencies decrease faster than what is implied by the empirical relations obtained for the typical type III bursts \citep[e.g.,][]{alv73}, meaning that the broadband bursts in the considered event look more like J-bursts or hockey sticks in the dynamic spectrum (Figure \ref{DSlarge}).

\subsection{Dynamics of the emission sources}
To determine the source size and position, we fitted the LOFAR data at each time and frequency by an elliptical Gaussian; the resulting source centroid position can be determined with an accuracy much better than the instrument beam width \citep[see, e.g.,][]{kon17}. Typical temporal evolution of the DPB source parameters at a single frequency is shown in Figure \ref{DSsmall}b.  We believe that the observed variations are related mainly to switching between the DPB sources and the background emission sources (whose parameters are less reliably determined).

\begin{figure*}
\includegraphics{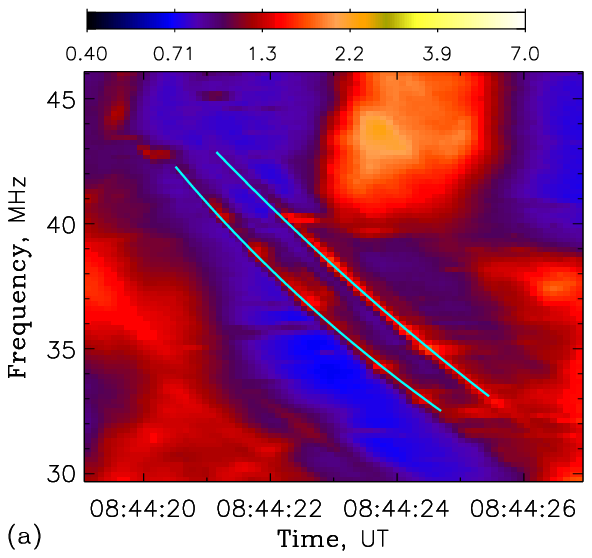}%
\includegraphics{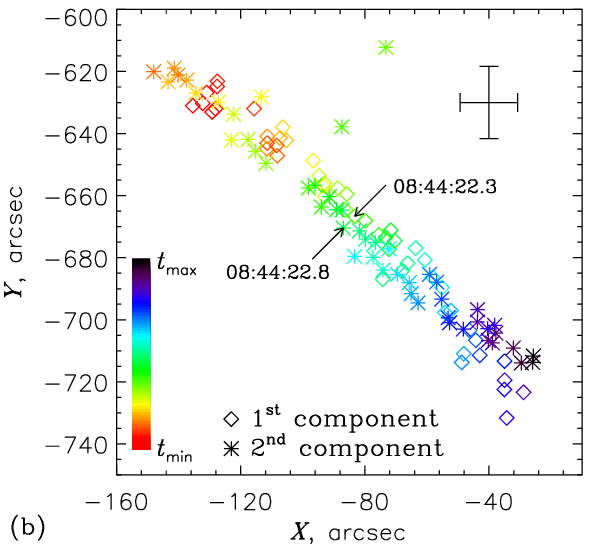}%
\includegraphics{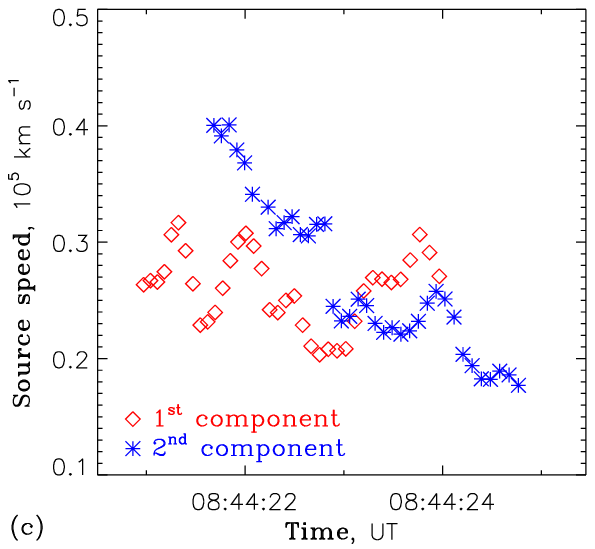}
\caption{Characteristics of a typical DPB with negative frequency drift. a) Dynamic spectrum (with approximate axial lines of the drifting components). b) Centroid locations of the emission sources at different times along the burst components (colors from red to violet correspond to increasing time); the times corresponding to nearby locations of the sources of different components highlight the delay between the components. The cross in the upper-right corner shows average error bars of the centroid locations (at $1\sigma$ level). c) Source speed (in the image plane) estimated using linear fits over 15 consecutive points.}
\label{srcA}
\end{figure*}

\begin{figure*}
\includegraphics{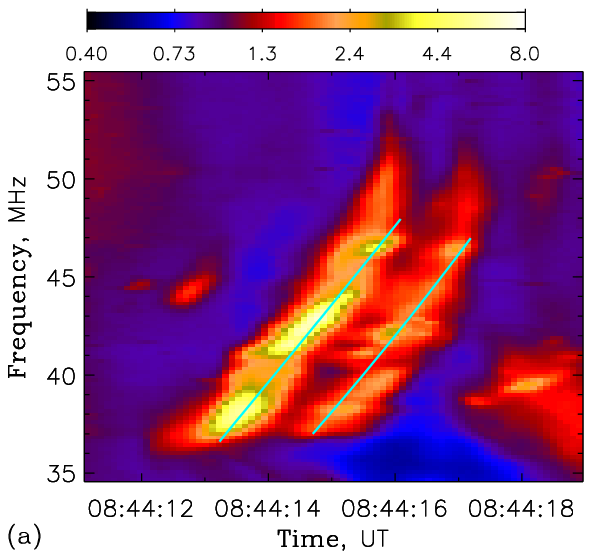}%
\includegraphics{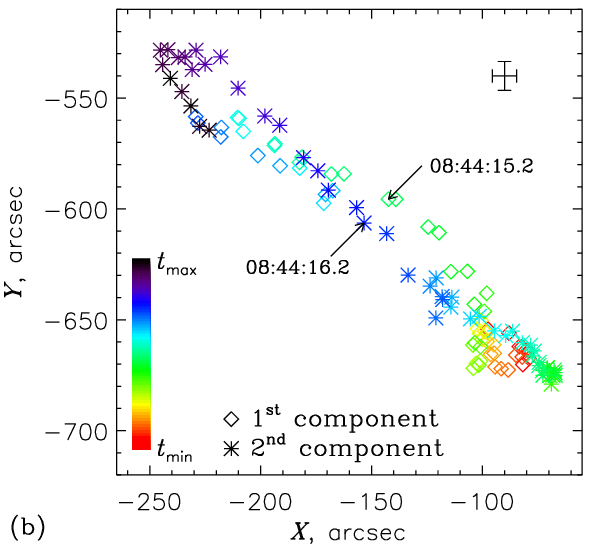}%
\includegraphics{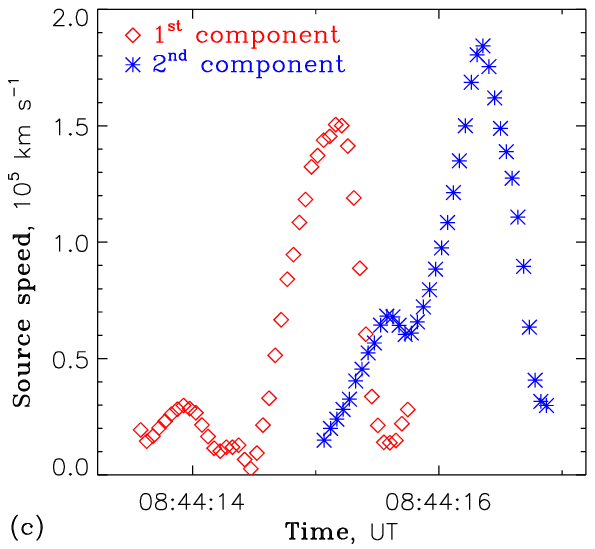}
\caption{Same as in Figure \protect\ref{srcA}, but for a DPB with positive frequency drift.}
\label{srcB}
\end{figure*}

\begin{figure*}
\includegraphics{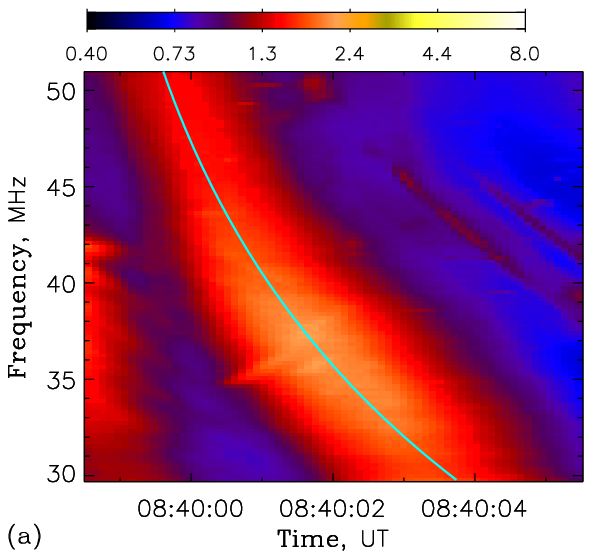}%
\includegraphics{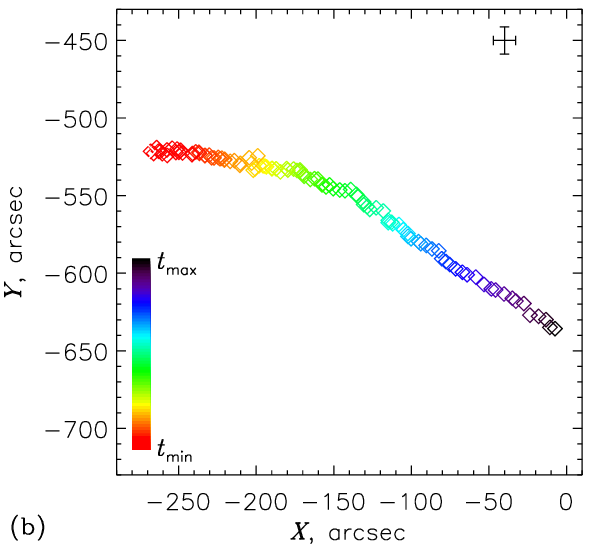}%
\includegraphics{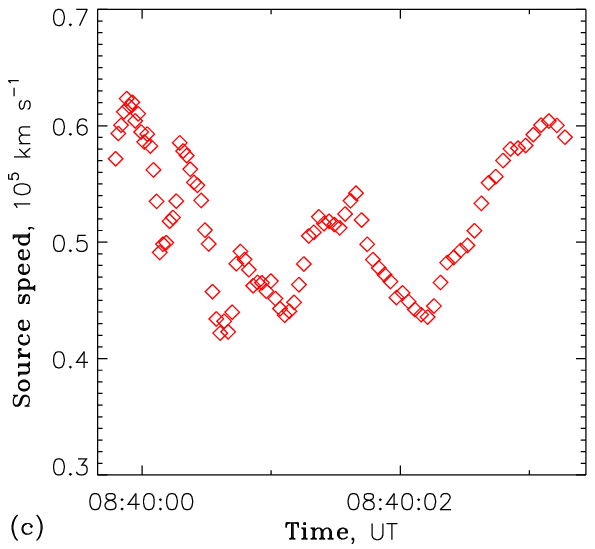}
\caption{Same as in Figure \protect\ref{srcA}, but for a broadband J-like burst.}
\label{srcS}
\end{figure*}

Figures \ref{srcA}--\ref{srcS} show the spatial evolution of the source centroids in time ($t$) and frequency ($f$); namely, the centroid coordinates follow the ``axial lines'' of the bursts determined as polynomial fits to the dependencies of $\log f$ versus $t$ describing the frequency drift of the bursts. In particular, Figure \ref{srcA} shows the analysis results for a typical DPB with negative frequency drift, with the dynamic spectrum in Figure \ref{srcA}a. Figure \ref{srcA}b shows the centroid locations of the emission sources at different times (two components of the drift pair are shown by different symbols, and the times are color-coded). The centroids exhibit an evident motion in the southeast direction. The sources of both components of the drift pair propagate in the same direction along nearly the same trajectory, although with a certain delay ($\sim 0.5-1$ s), as indicated by the time labels. Figure \ref{srcA}c shows the visible source speed (estimated using linear fits over 15 consecutive points); this speed varies from $\sim 20\,000$ to $\sim 40\,000$ km $\textrm{s}^{-1}$.

Figure \ref{srcB} shows the similar analysis for a DPB with positive frequency drift. Again, the sources of both components of the drift pair propagate in the same direction along nearly the same trajectories with a delay of $\sim 1-1.5$ s (Figure \ref{srcB}b). On the other hand, the source trajectories are now somewhat more complicated: although the sources move predominantly in the northwest direction (i.e., opposite to that for the negatively drifting burst), one can see more complex motions with direction reversals at the beginning and the end of the emission stripes. These features are likely related to the projection effects: even if the actual three-dimensional trajectory of the emission source is directed steadily downwards, its projection on the image plane can exhibit loops and reversals. For the same reason, the visible (projected) source velocity varies in a broad range -- from very low absolute values up to $\sim 180\,000$ km $\textrm{s}^{-1}$ (Figure \ref{srcB}c), which is among the fastest motions detected in the considered event and much higher than the speed inferred from the frequency drift rate. Such high speeds are likely to be associated with apparent motion of the scattered image rather than the actual motion of the emitter \citep[cf.][]{kon17, sha18}.

Finally, Figure \ref{srcS} shows the spatial evolution of the source of a broadband drifting burst\footnote{The analysis was restricted to the frequencies below 50 MHz.}, which proved to be qualitatively similar to that of the DPBs. Namely, the emission source moves in an approximately southeast direction; the source velocity in the image plane is of about $50\,000$ km $\textrm{s}^{-1}$. This velocity is again consistent with radio-wave scattering.

\begin{figure}
\includegraphics{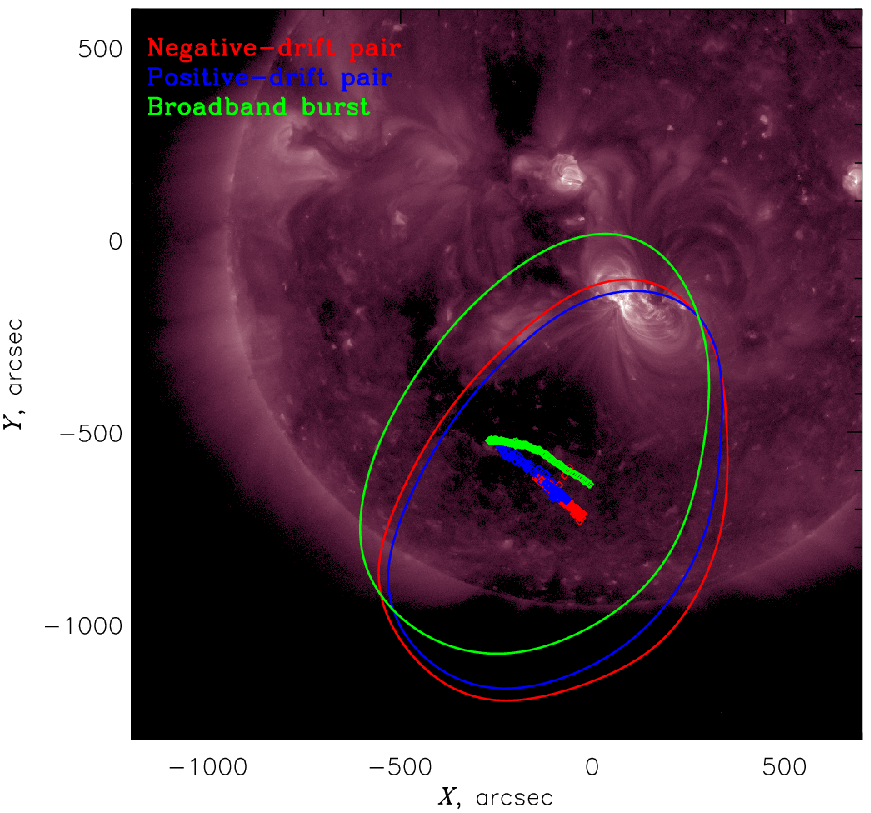}
\caption{Centroid locations of the emission sources of three selected emission features (same as in Figures \protect\ref{srcA}--\ref{srcS}) overplotted on the SDO AIA 211 {\AA} image. Contours of the radio intensity maps (corresponding to the frequency of 38 MHz) at half-maximum level are also shown.}
\label{over}
\end{figure}

In Figure \ref{over}, the centroid locations of the emission sources of the three above-mentioned bursts are plotted in one image. The sources of the DPBs with the positive and negative frequency drifts coincide spatially (where their frequency ranges overlap). The broadband J-like burst originates from the same region; a similar conclusion was made earlier by \citet{suz79}. The radio emission sources demonstrate a possible association with a coronal hole and an active region and therefore the emission may be related either to processes in the hole or to interaction between the hole and the nearby active region. We note, however, that {absolute} position of solar radio sources is less constrained because it can be affected by the ionosphere \citep[see, e.g.][]{gor19}. The source height above photosphere is about $3/4{R_{\sun}}$ assuming plasma emission, and therefore the projection effect is an additional complicating factor.

Analysis of other radio bursts in the considered event has revealed qualitatively similar features. The source centroid trajectories can be rather complicated (including self-intersections); this behavior was observed in both the DPBs and the broadband J-like bursts. Despite the complex motion, the emission sources of the components of the same DPBs propagate in the same direction along nearly the same trajectories (with the deviations not exceeding a few tens of arcseconds). The bursts occurring at similar times and frequencies (including both the DPBs with opposite frequency drifts and the broadband bursts) originate from the same regions. The visible source velocities of the DPBs vary from about $20\,000$ to $\gtrsim 100\,000$ km $\textrm{s}^{-1}$.

\begin{figure}
\includegraphics{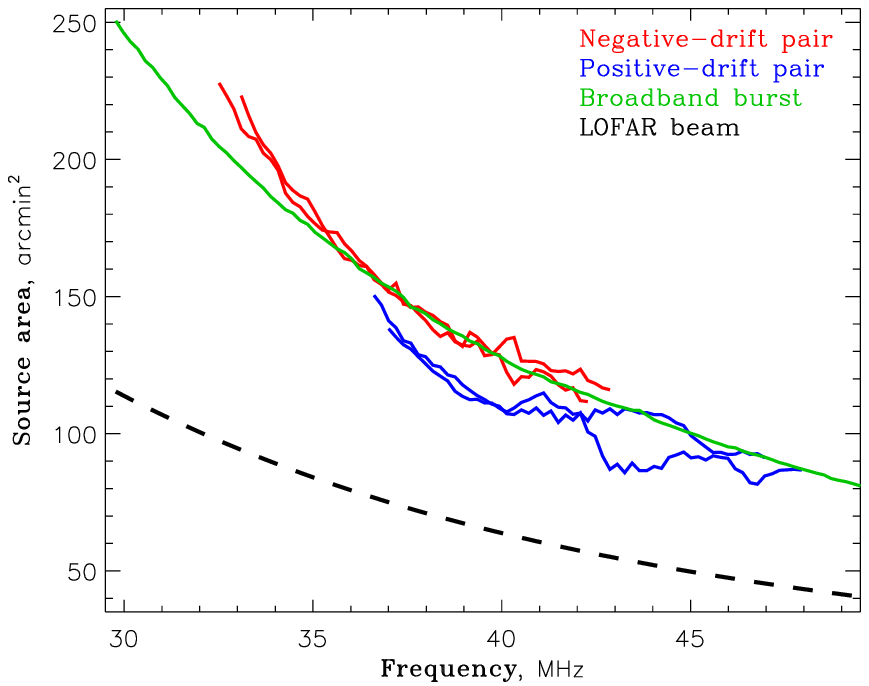}
\caption{Estimated areas of the emission sources of three selected emission features (same as in Figures \protect\ref{srcA}--\protect\ref{srcS}) at half-maximum level. The LOFAR beam area is plotted by the dashed line.}
\label{area}
\end{figure}

\subsection{Emission source sizes}\label{ssize}
Figure \ref{area} shows the (visible) areas of the emission sources of two DPBs and one broadband burst (the same as in Figures \ref{srcA}--\ref{srcS}) estimated using the Gaussian fitting. The source areas decrease with frequency approximately as $\propto f^{-2}$, from $\sim 250$ $\textrm{arcmin}^2$ at 30 MHz to $\sim 80$ $\textrm{arcmin}^2$ at 50 MHz; they slightly exceed  (by a factor of $\sim 1.5-2.3$) the instrument beam area. Notably, both the DPBs and the broadband J-like bursts have nearly the same source sizes.

The above result is intriguing because the emission sources of type III bursts at the considered frequencies are much larger, being of $\sim 450$ $\textrm{arcmin}^2$ at 32 MHz \citep{kon17} and $\sim 320$ $\textrm{arcmin}^2$ at 43 MHz \citep{dul80}. On the other hand, our observations agree with the earlier results of \citet{suz79}, who found that the sources of the DPBs and the associated broadband bursts have the same size ($\sim 120-160$ $\textrm{arcmin}^2$ at 43 MHz), and that the broadband bursts associated with DPBs have  considerably smaller sources than those of the typical type III bursts \citep[e.g.,][]{dul80}.

\section{Discussion}
\citet{rob58} attributed the formation of the DPBs to the radio echo effect. This model was later rejected on the basis that it predicts a significant difference in the visible source locations of the first (direct) and the second (reflected) components of a drift pair, which is not actually observed \citep[e.g.,][]{suz79}. Our observations show for the first time that the source locations of the pair components have similar motions and size variations at a given frequency. From the peak to decay, the sources of both components increase in size and show radial motion in time in a repeating manner. A similar pattern (but without pairing component) is observed in type IIIb bursts \citep{kon17}. The additional unique observational aspect is that the speed of the source locations can greatly exceed the speed of the exciter as determined from the dynamic spectra, which further emphasizes the importance of radio-scattering effects. The similarity of source motions and size expansions suggest the dominance of radio propagation effects. Thus, since the scattering of radio waves changes the apparent source positions, both components will be located at the distance of the last-scattering surface \citep{kon19} rather than near the emission and reflection locations. Therefore, we conclude \citep[contrary to the previous works by][]{rob58, suz79} that despite the similar positions of the components, the drift pairs are consistent with scattering. Since the scattering will also be present for radio waves propagating downwards, the delay will be a function of scattering. Therefore, since the propagation (scattering) effects seem to be a dominant factor determining the visible source location at metric wavelengths, we believe that turbulent-scattering echo is a likely scenario and requires further investigation.

\citet{mol78} proposed the model in which the components of a DPB are generated by two magnetohydrodynamic shocks propagating (with some relative delay) in opposite directions from the axis of a coronal streamer; the similarity of the pair components was explained by the highly symmetric structure of a streamer. However, this model implies that the sources of the pair components should propagate in opposite directions, while in our observations they always propagate in the same direction.

\citet{zai78} proposed the model based on the double plasma resonance (DPR) effect (similar to the commonly accepted model for zebra patterns). However, in this model the pair components should be shifted in frequency rather than in time; also, there should be a certain systematic shift between the source locations of the pair components at the same frequency. The latter prediction is not ruled out by our observations because the expected shift can be relatively small, and the observations are affected by projection effects. Nevertheless, this model would require a very specific configuration of magnetic field and plasma, with the DPR levels running in parallel for $\gtrsim 2\times 10^5$~km (the observed length of the source paths) and an emitting agent propagating with a velocity of up to $\sim 10^5$ km $\textrm{s}^{-1}$ {along} these levels, which exceeds the drift-rate speed of $\sim 10^4$~ km/s.

In summary, the first imaging spectroscopy observations do not support counter-propagating shocks or models with pure frequency shift of the components. The key factors for identifying the formation mechanism of DPBs are likely to be parallel motion of the component sources, shift in time rather than in frequency, nearly identical time variation of size and position for two components at a single frequency, and fast motion of their apparent sources. The latter two factors are reminiscent of the fundamental type IIIb sources \citep{kon17}, where scattering of the waves on plasma turbulence increases the visible source size and is responsible for high apparent source velocities \citep{sha18}. Besides affecting the apparent source size and position, anisotropic scattering would make the emission time profiles similar for both components, thus contributing to the formation of the characteristic dynamic spectra of the DPBs. A likely interpretation is therefore that the DPBs are produced at fundamental plasma frequency  while their delayed second component is due to scattering in turbulent media, probably with higher anisotropy than in the type III source regions, which produces a less diffuse echo \citep{kon19}. Anisotropic scattering implies that the probability of detecting a DPB should be dependent on the viewing angle.

\begin{acknowledgements}
AAK acknowledges partial support from the RFBR grant 17-52-10001, budgetary funding of Basic Research program II.16, and the program KP19-270 of the RAS Presidium. EPK was supported from STFC grant ST/P000533/1.
This paper is based (in part) on data obtained from facilities of the International LOFAR Telescope (ILT) under project code LC8\_027. LOFAR \citep{haa13} is the Low Frequency Array designed and constructed by ASTRON. It has observing, data processing, and data storage facilities in several countries, which are owned by various parties (each with their own funding sources), and that are collectively operated by the ILT foundation under a joint scientific policy. The ILT resources have benefited from the following recent major funding sources: CNRS-INSU, Observatoire de Paris and Universit\'e d'Orl\'eans, France; BMBF, MIWF-NRW, MPG, Germany; Science Foundation Ireland (SFI), Department of Business, Enterprise and Innovation (DBEI), Ireland; NWO, The Netherlands; The Science and Technology Facilities Council, UK. The authors acknowledge the support by the international team grant (\url{http://www.issibern.ch/teams/lofar/}) from ISSI Bern, Switzerland.
\end{acknowledgements}

\end{document}